\documentclass[fleqn,usenatbib]{mnras}

\usepackage{newtxtext,newtxmath}

\usepackage[T1]{fontenc}

\DeclareRobustCommand{\VAN}[3]{#2}
\let\VANthebibliography\thebibliography
\def\thebibliography{\DeclareRobustCommand{\VAN}[3]{##3}\VANthebibliography}

\usepackage{graphicx}	
\usepackage{amsmath}	
\usepackage{color}
\usepackage{float}
\usepackage{subfigure}
\usepackage{enumerate}
\usepackage{multirow}
\usepackage{booktabs}
\usepackage{threeparttable}
\graphicspath{{./}{figures/}}

\title[Multi-wavelength study of SNR G51.26+0.11]{A Study of GeV Gamma-ray Emission toward Supernova Remnant G51.26+0.11 and Its Molecular Environment}

\author[W. J. Zhong et al.]{
Wen-Juan Zhong,$^{1}$
Xiao Zhang,$^{1,2}$\thanks{E-mail: xiaozhang@nju.edu.cn}
Yang Chen$^{1,2}$\thanks{E-mail: ygchen@nju.edu.cn}
and Qian-Qian Zhang$^{1}$
\\
$^{1}$School of Astronomy \& Space Science, Nanjing University, Nanjing 210023, China\\
$^{2}$Key Laboratory of Modern Astronomy and Astrophysics,
Nanjing University, Ministry of Education, 
Nanjing 210023, China
}

\date{Accepted 2023 February 23. Received 2023 February 19; in original form 2022 October 5}

\pubyear{2023}

\begin{document}
\label{firstpage}
\pagerange{\pageref{firstpage}--\pageref{lastpage}}
\maketitle

\begin{abstract}
We reanalyze the \emph{Fermi}-LAT GeV $\gamma$-ray emission in the region of supernova remnant (SNR) G51.26+0.11 and investigate its interstellar molecular environment with the CO-line data.
At GeV energies, based on 13.2 years of 
\emph{Fermi}-LAT data, the extended $\gamma$-ray emission observed in this region is resolved into a uniform-disk source (`Src A') with a significance of 19.5$\sigma$ and a point source (4FGL J1924.3+1628) with a significance of 4.2$\sigma$ in 0.2--500 GeV.
With an angular radius of $\sim$ 0.17$^\circ$, `Src A' overlaps with SNR G51.26+0.11 significantly in the line of sight.
On the other hand, the morphological coincidence between the SNR and the $\sim +54$ km s$^{-1}$ molecular clouds (MCs) together with the asymmetric or broad $^{12}$CO line profiles near the SNR boundary provides evidence for the very likely SNR-MC interaction.
The SNR-MC interaction and the H\,{\small I} absorption features indicate that SNR G51.26+0.11 is located at a kinematic distance of $6.2\pm0.5$ kpc.
Combined with the results from the multi-wavelength analysis, the $\gamma$-ray emission of the SNR (`Src A') can be naturally explained by a hadronic model with a soft power-law proton spectrum of index $\sim$ 2.25.
\end{abstract}




\section{Introduction}\label{sec:intro}

\begin{figure}
    \includegraphics[width=\columnwidth]{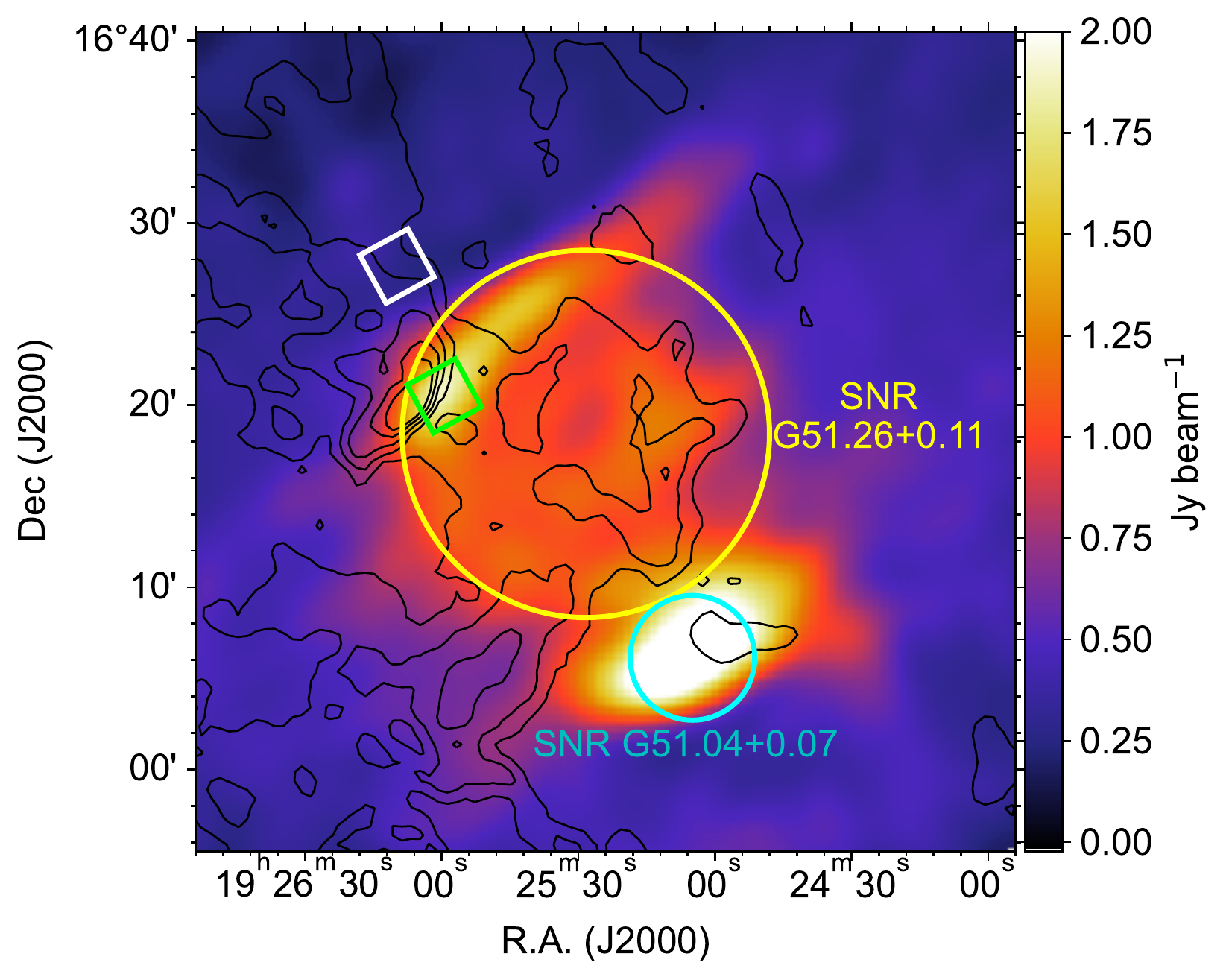} 
        \caption{
        SNR G51.26+0.11 and SNR G51.04+0.07 as seen in the GLEAM 200 MHz data \citep{hurley19}, superposed with MWISP $^{12}$CO ($J=1$ -- $0$) contours (black) in levels of 15, 25, 35, 45 and 55 K km s$^{-1}$ in the velocity range of +52 -- +58 km s$^{-1}$.
        The yellow and cyan circles, respectively, depict the approximate radio boundaries of the SNRs, according to \citet{dokara18}.
        The white and green boxes denote the regions from which the H\,{\small I} spectra of `Background' and `SNR' are extracted.
        }
  \label{fig:radio}
\end{figure}

The acceleration of cosmic rays (CRs) can be explored through observations of the $\gamma$-ray emission resulting from the decay of $\pi^{0}$ mesons produced via proton-proton collisions \citep{acker13}.
Supernova remnants (SNRs) are a sort of the popular candidates for Galactic CR accelerators by virtue of their strong shocks, while molecular clouds (MCs) can provide target gas for proton-proton hadronic interaction.
Therefore, studies of the $\gamma$-ray emission from SNR-MC associations are crucial to search for the signatures of proton acceleration, and multi-wavelength observations are necessary to reveal their interstellar environments.
Dozens of GeV $\gamma$-ray sources associated with SNRs have been discovered with the Large Area Telescope (LAT) on board the \emph{Fermi Gamma-ray Space Telescope}, among which SNRs interacting with nearby MCs appear to be brighter and softer \citep{liu15,acero16}.
\\
\indent G51.21+0.11 was classified as an SNR candidate by \citet{anderson17} using the combination of 1--2 GHz continuum data from The HI, OH, Recombination line survey of the Milky Way (THOR) and 1.4 GHz Very Large Array Galactic Plane Survey continuum data and observed by \citet{driessen18} using the LOw Frequency ARray.
Following studies identified it as a complex of two separate SNRs: the compact SNR G51.04+0.07 \citep{supan18} and the shell-type SNR G51.26+0.11 \citep{dokara18}.
Evidence of nonthermal emission from the two distinct regions in this complex was also provided \citep{supan18,dokara18}. 
Figure \ref{fig:radio} displays the radio emission from the complex in the 80--300 MHz GaLactic and Extragalactic All-sky Murchison Widefield Array survey (GLEAM; \citealp{wayth15,hurley17}) data.
\\
\indent The SNR nature of G51.26+0.11 was further confirmed by \citet{dokara21} using the data of the GLObal view of STAR formation in the Milky Way survey that was conducted with the \emph{Karl G. Jansky} VLA.
They measured a degree of polarization of $0.06\pm0.02$ and a radio spectral index of $\sim -0.4$, which are consistent with the case of SNRs.
The measured 200 MHz and 1.4 GHz flux densities are $25.8\pm 3.6$ Jy and $12.4\pm 0.6$ Jy, respectively.
Recently, \citet{ranas22} estimated the distance of $6.6\pm 1.7$ kpc to G51.26+0.11 using the H\,{\small I} absorption spectra constructed with the THOR data.
At GeV energies, \citet{araya21} performed a study of the $\gamma$-ray emission with 12.3 years of \emph{Fermi}-LAT data and found an extended $\gamma$-ray source with a significance of 21$\sigma$ above 200 MeV in the region of G51.26+0.11.
They suggested that the spectral energy distribution (SED) observed could be fitted by hadronic or inverse Compton emission.
In the hadronic scenario, the spectral index of the protons is $\sim$ 2.18 and G51.26+0.11 could be the cause of the $\gamma$-rays.
The MCs at the local-standard-of-rest (LSR) velocities +43.4 -- +52.2 and +53 -- +58.8 km s$^{-1}$ were suggested to be compatible with the hadronic scenario. 
Meanwhile, the pulsar wind nebula (PWN) origin could not be ruled out or confirmed.
\\
\indent In this study, we reanalyze the GeV emission from SNR G51.26+0.11 using \emph{Fermi}-LAT data and investigate the interstellar molecular environment of the SNR with millimeter CO-line observations.
We focus on the morphology of the $\gamma$-ray emission and the possible hadronic contribution resulting from the SNR-MC interaction.
In Section \ref{sec:data}, we describe the $\gamma$-ray and multi-wavelength observations and data reduction.
In Section \ref{sec:multi}, the corresponding analysis and results are presented.
The results are discussed in Section \ref{sec:discuss} and summarized in Section \ref{sec:sum}.

\section{Observations and Data} \label{sec:data}

\subsection{\emph{Fermi}-LAT Observational Data} \label{subsec:fermidata}

The Large Area Telescope (LAT) onboard the \emph{Fermi Gamma-ray Space Telescope} surveys the $\gamma$-ray sky in the 20 MeV to more than 300 GeV energy range.
We collect more than 13.2 years of \emph{Fermi}-LAT Pass 8 data toward a circular region which is $15^{\circ}$ in radius and centered at the coordinates R.A. = $291.3^\circ$, Dec = $16.3^\circ$ (J2000),
from 2008-08-04 15:43:36 (UTC) to 2021-11-07 02:26:50 (UTC). 
We analyze the data with the standard softwares {\small Fermitools}\footnote{\url{http://fermi.gsfc.nasa.gov/ssc/data/analysis/software/}}
version 2.0.8 released on 2021 January 20, and {\small Fermipy}\footnote{\url{https://fermipy.readthedocs.io/en/stable/}} version 1.0.1 released on 2021 March 12. 
We select `SOURCE' class and `FRONT+BACK' type events (evclass = 128, evtype = 3) with zenith angle $<90^\circ$ to eliminate Earth limb events, and restrict the energy range to 0.2--500 GeV.
To choose good time intervals, we apply the recommended filter string `(DATA\_QUAL==1)\&\&(LAT\_CONFIG==1)' in tool \texttt{gtmktime}.
For `SOURCE' events, the instrument response function (IRF) `P8R3\_SOURCE\_V3\_v1' is used.
To build the background model, we include the LAT sources listed in the \emph{Fermi}-LAT Fourth Source Catalog Data Release 2 (4FGL-DR2, \citealt{ballet20}) in a radius of $25^\circ$ around the centre of the region of interest (ROI), as well as the Galactic diffuse emission (\emph{gll\_iem\_v07.fits}) and isotropic emission (\emph{iso\_P8R3\_SOURCE\_V3\_v1.txt}). 

\subsection{CO Line Data} \label{subsec:codata}

We make use of the data from the Milky Way Imaging Scroll Painting (MWISP) survey project.
The CO observation was made in the $^{12}$CO ($J=1$ -- $0$) line (at 115.271 GHz), and the $^{13}$CO ($J=1$ -- $0$) line (at 110.201 GHz) in 2016 June using the 13.7 m millimeter-wavelength telescope of the Purple Mountain Observatory at Delingha, China.
We focus on a $1.5^\circ\times1.5^\circ$ area covering G51.26+0.11 centred at $l=51.26^\circ$, $b=+0.11^\circ$ in the Galactic coordinate system with a grid spacing $\sim30^{\prime\prime}$.
The half-power beam width of the telescope is about 50$^{\prime\prime}$ for the two lines.
The mean RMS noise level of the main beam brightness temperature is $\sim$ 0.52 K for the $^{12}$CO ($J=1$ -- $0$) at the velocity resolution of 0.158 km s$^{-1}$ and $\sim$ 0.25 K for the $^{13}$CO ($J=1$ -- $0$) lines at the velocity resolution of 0.166 km s$^{-1}$.

\subsection{Other Data} \label{subsec:h1data}
As an aid to constraining the distance to SNR G51.26+0.11, we obtain the 1420 MHz radio continuum and H\,{\small I} line emission data from the Very Large Array Galactic Plane Survey \citep{vla06}.
The continuum image has a spatial resolution of $18^{\prime\prime}$.
The synthesized beam for the H\,{\small I} spectral line images is $18^{\prime\prime}$ and the radial velocity resolution is 0.824 km s$^{-1}$. 

\section{Multi-wavelength Analysis} \label{sec:multi}

\subsection{\emph{Fermi}-LAT Gamma-ray Emission} \label{subsec:fermi}

\subsubsection{Morphological Analysis} \label{subsec:morph}
\begin{figure*}
    \centering
    \includegraphics[width=1.0\textwidth]{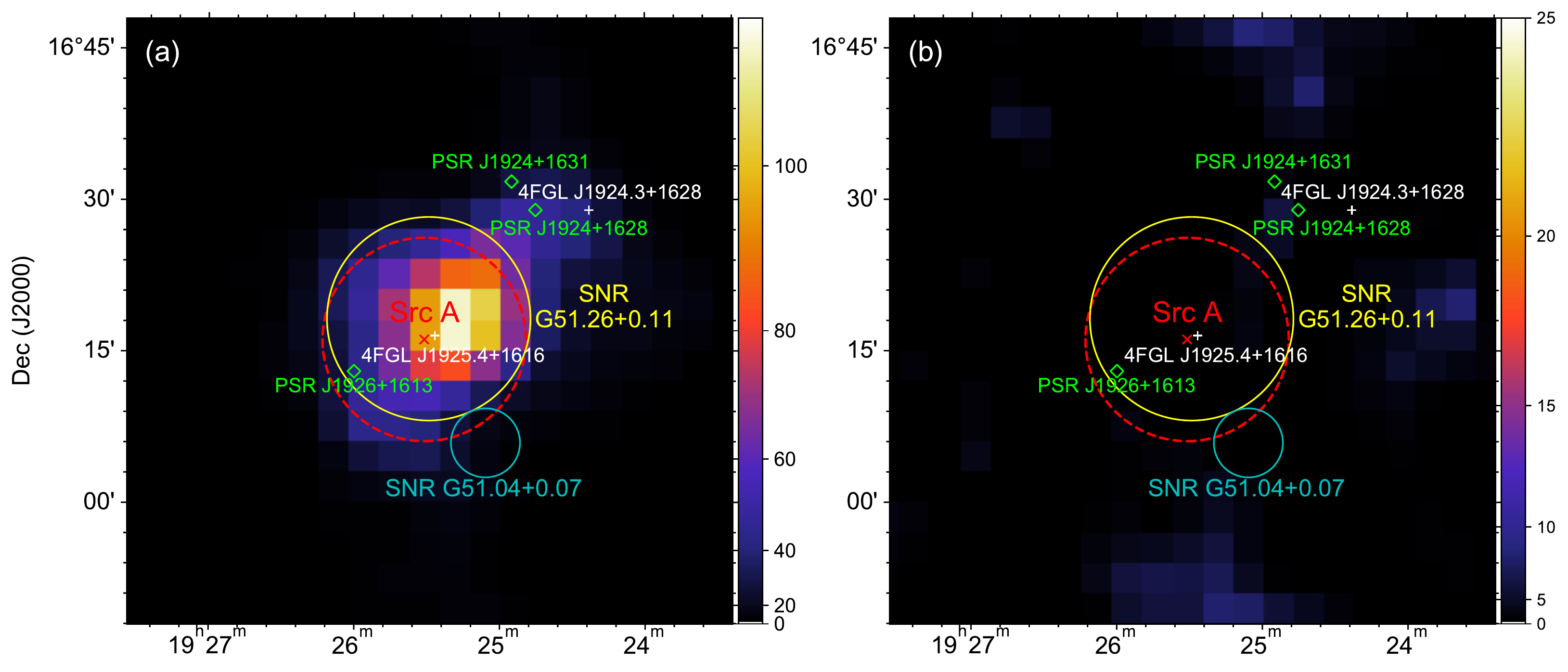} 
    \includegraphics[width=1.0\textwidth]{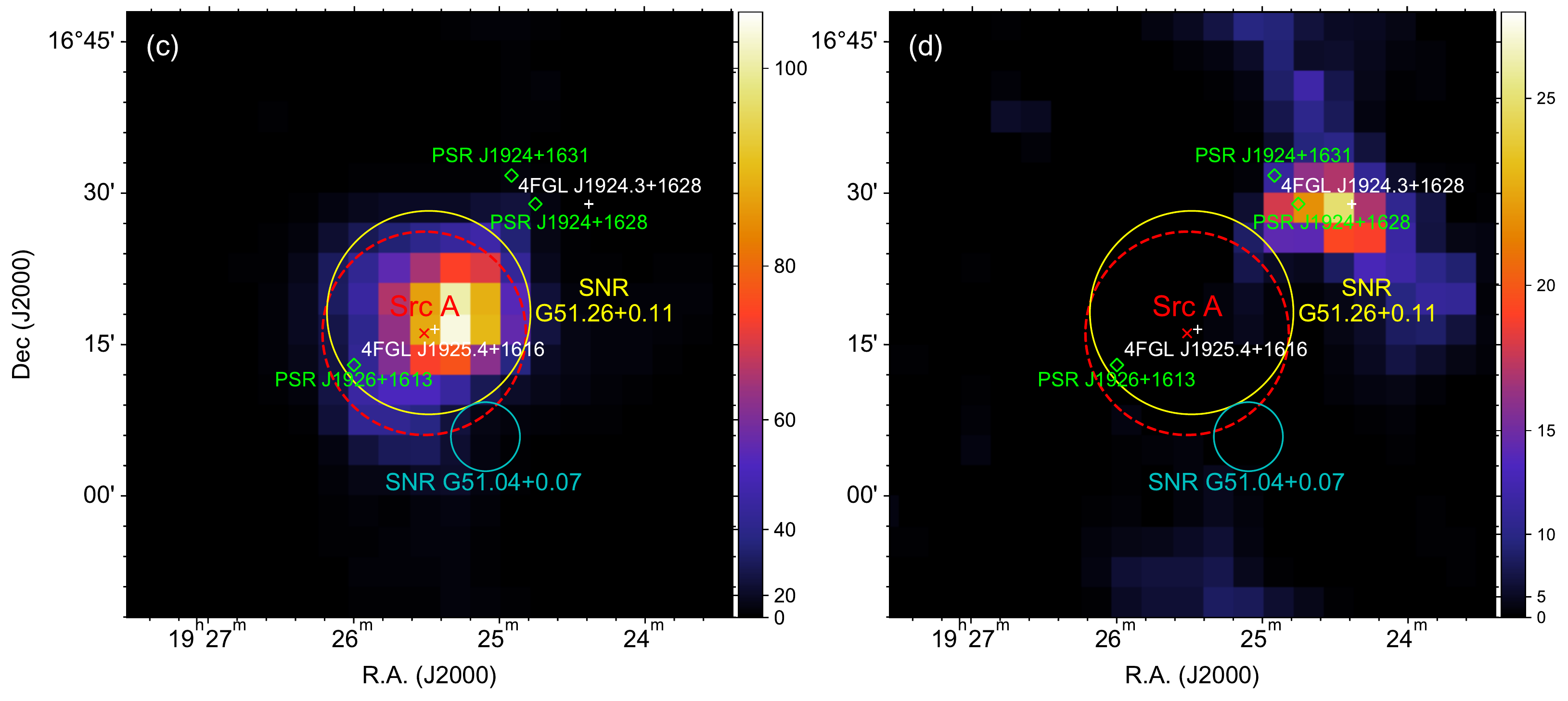}
        \caption{
        Background-subtracted TS maps of the $1^\circ \times 1^\circ$ region centred at SNR G51.26+0.11 in the energy range 5--500 GeV.
        The image scale of the maps is $0.05^\circ$ per pixel.
        The yellow and cyan circles are the same as those in Figure \ref{fig:radio}.
        The red cross indicates the best-fitting position of the extended source (`Src A') and the dashed red circle depicts the 68$\%$-containment region of the disk template. 
        The white crosses show the positions of two 4FGL-DR2 catalogue sources: 4FGL J1925.4+1616 and 4FGL J1924.3+1628.
        The green diamond signs represent the position of three pulsars in this region: PSR J1926+1613, PSR J1924+1631 and PSR J1924+1628.
        (a) 4FGL J1925.4+1616 and 4FGL J1924.3+1628 are excluded from the source model.
        (b) `Src A' and 4FGL J1924.3+1628 are included in the source model.
        (c) `Src A' is excluded from the adjusted source model. 
        (d) 4FGL J1924.3+1628 is excluded from the adjusted source model.
        }
  \label{fig:tsmap}
\end{figure*}

\begin{table*}
  \centering
  \caption{Best-fitting parameters in different spatial models ($>$5 GeV).}
    \begin{tabular}{ccccccc}
    \toprule
    \toprule
    \multicolumn{2}{c}{Morphological model} & R.A.(J2000) $^a$ & Dec(J2000) $^a$ & Extension ($\sigma$) $^b$ & $\Delta k$ $^c$ & $\Delta$AIC $^c$ \\
      &   & ($^{\circ}$) & ($^{\circ}$) & ($^{\circ}$) &  &  \\
    \midrule
    Single-source & \emph{1Gauss} $^d$ & 291.31$\pm 0.02$ & 16.30$\pm 0.02$ & 0.22$^{+0.04}_{-0.03}$ & 0 & 0 \\
    \midrule
    \multirow{4}[4]{*}{Two-source} & \emph{1Gauss} & 291.35$\pm 0.02$ & 16.28$\pm 0.02$ & 0.18$^{+0.04}_{-0.03}$ & \multirow{2}[2]{*}{2} & \multirow{2}[2]{*}{-7.87} \\
      & \emph{1PS} $^e$ & 291.10 & 16.48 & $-$ &   &  \\
\cmidrule{2-7}      & \emph{1Disk} & 291.38$\pm 0.02$ & 16.27$\pm 0.02$ & 0.17$^{+0.02}_{-0.02}$ & \multirow{2}[2]{*}{2} & \multirow{2}[2]{*}{-12.49} \\
      & \emph{1PS} $^e$ & 291.10 & 16.48 & $-$ &   &  \\
    \bottomrule
    \end{tabular}%
    \begin{tablenotes}
		\footnotesize
		\item \textbf{Notes.}
		\item $^a$ The fitted positions with 1$\sigma$ statistical uncertainty.
		\item $^b$ The respective 68$\%$-containment radii.
		\item $^c$ $k$ and AIC values are provided as differences with respect to the \emph{1Gauss} model.
		\item $^d$ The values of position and extension are from \citet{araya21}.
        \item $^e$ The position of the point source is fixed to be the same as that of the catalogued source 4FGL J1924.3+1628.
	\end{tablenotes}
  \label{tab:morph_aic}%
\end{table*}%

There are two 4FGL-DR2 catalogue sources, 4FGL J1925.4+1616 and 4FGL J1924.3+1628, toward SNR G51.26+0.11.
To analyze the morphology of the GeV emission in the ROI, we remove these two sources from the source models.
To reduce uncertainties caused by large point-spread function at lower energies, we select photon events at energies above 5 GeV in this region.
Then, we perform the binned maximum likelihood analysis and apply the energy dispersion correction.
We free the normalization parameters of the Galactic and isotropic diffuse background components and the sources within $10^\circ$ from the ROI centre, as well as the spectral parameters of the sources within $3^\circ$ from the ROI centre while fitting the models.
Finally, we generate the residual test-statistic (TS) map of the $1.5^\circ \times 1.5^\circ$ region centred at G51.26+0.11 in the energy range of 5--500 GeV. 
The test statistic is defined as ${\rm TS} = -2\log (\mathcal{L}_0 / \mathcal{L}_1)$, in which $\mathcal{L}_0$ is the likelihood of the null hypothesis and $\mathcal{L}_1$ is the likelihood of the hypothesis being tested.
The detection significance $\sigma$ is usually approximated by the square root of the TS \citep{wilks38}.
As shown in Figure \ref{fig:tsmap}(a), 
the excess of the TS map shows an elongated shape toward the northwest, suggesting that an extended symmetric Gaussian may not be sufficient to fit well.
\\
\indent We define the following three source models to describe the $\gamma$-ray emission, where the notation \emph{PS} stands for point sources, \emph{Gauss} for the Gaussian template, and \emph{Disk} for the disk template with a uniform model.
\begin{enumerate}[i.]
\item \emph{1Gauss} includes only an extended source with a Gaussian model assuming a power-law (PL)-type spectrum;
\item \emph{1Gauss+1PS} includes an extended source with a Gaussian model and a point source, which are assumed to have PL-type spectra;
\item \emph{1Disk+1PS} includes an extended source with a uniform disk model and a point source, which are assumed to have PL-type spectra.
\end{enumerate}

We place the extended source and the point source mentioned above (called target sources) at the same positions as the catalogued sources 4FGL J1925.4+1616 and 4FGL J1924.3+1628, respectively.
For the models including the point source (4FGL J1924.3+1628), 
we fix the location of the point source to the original coordinates of 4FGL J1924.3+1628.
For all three models, we use method \texttt{extension} of {\small Fermipy} to obtain the best-fitting extension and position of the extended source (4FGL J1925.4+1616).
We also calculate the likelihood ratio between the best-fitting extended model and the point source hypothesis, defined as TS$_{\rm ext}$ $\equiv 2 \log ({\mathcal{L}_{\rm ext}} / {\mathcal{L}_{\rm PS}})$.
The $\gamma$-ray source is considered to be significantly extended only if its TS$_{\rm ext}$ exceeds 16 \citep{lande12}.
In all three models, the measurement results indicate that the extended nature (of 4FGL J1925.4+1616) is genuine (TS$_{\rm ext}$=48.4, 32.2, and 29.3, respectively).
In the \emph{1Gauss} model, the best-fitting position and 68\%-containment radius are R.A. = $291.31\pm 0.02^{\circ}$, Dec = $16.30\pm 0.02^{\circ}$ (J2000) and $\sigma={0.24^{+0.04}_{-0.03}}^{\circ}$, respectively, which are consistent with those reported by \citet{araya21}.
For comparison, we replace the spatial parameters of the \emph{1Gauss} model with Araya's results in the following analysis.
The best-fitting spatial parameters of the three models are summarized in Table \ref{tab:morph_aic}.
\\
\indent We add each of the three templates to the background model and repeat the fit with {\small Fermitools}.
During the fitting process, we only free the normalization and spectral parameters of the target sources as well as the normalization of the Galactic diffuse emission, while all the other parameters are fixed to their best-fitting values from the above analysis.
We use the Akaike Information Criterion (AIC; \citealt{lande12}) to select the best-fitting spatial model.
AIC is defined as ${\rm AIC}=2k-2\ln \mathcal{L}$, 
where $k$ is the number of free parameters in the model and $\mathcal{L}$ is the likelihood of the model tested in the analysis.
The model which minimizes the AIC is considered to be the best.
Compared with the \emph{1Gauss} model, the $\Delta$AIC values for the three models are tabulated in Table \ref{tab:morph_aic}.
As can be seen,
\emph{1Disk+1PS} could be the best-fitting spatial model, in which we consider 4FGL J1925.4+1616 as a uniform-disk source and 4FGL J1924.3+1628 as a point source.
With the inclusion of the \emph{1Disk+1PS} template in the background model, there is almost no residual $\gamma$-ray emission in the TS map (see Figure \ref{fig:tsmap}(b)).
Therefore,
we replace the template of 4FGL J1925.4+1616 with its best-fitting parameters and consider it as the GeV counterpart of G51.26+0.11, denoted as `Src A'.
As we include source 4FGL J1924.3+1628 in the source model, the angular radius of `Src A' ($\sim 0.17^{\circ}$) is a bit smaller than Araya's \citeyearpar{araya21} result.
The fitted TS values for `Src A' and 4FGL J1924.3+1628 in 5--500 GeV are 95.6 and 22.2, respectively, with the significance of 9.8$\sigma$ and 3.7$\sigma$.
The corresponding TS maps which exclude `Src A' and 4FGL J1924.3+1628 from the adjusted model are shown in Figures \ref{fig:tsmap}(c) and \ref{fig:tsmap}(d), respectively.

\subsubsection{Spectral Analysis} \label{subsec:spectral}

We select photon events in the energy range of 0.2--500 GeV for the spectral analysis.
After adjusting the background model with the \emph{1Disk+1PS} template, we free the normalization and spectral parameters of the sources within 5$^\circ$ from the ROI centre.
The normalization parameters of the Galactic and isotropic diffuse background components are also set free.
To study the spectral properties of `Src A', we test four spectral models: a power-law (PL), an exponentially cutoff power-law (PLEC), a log-parabola model (LogP), and a broken power-law (BPL).
The formulae and free parameters of these spectra are listed in Table \ref{tab:fml}.
Since the lowest $\Delta$AIC value points to the best-fitting spectral model, as shown in Table \ref{tab:spec_aic}, a PL spectrum is preferred for `Src A', and no significant curvature is detected.
\\
\indent Using the best-fitting PL spectral parameters for `Src A',
the obtained flux in 0.2--500 GeV is $2.86\times10^{-11}\ {\rm erg\ cm^{-2}\ s^{-1}}$ (with ${\rm \Gamma=2.19\pm 0.04}$), giving a luminosity of $7.3\times10^{34}\ {\rm erg\ s^{-1}}$ in 1--100 GeV at a distance of 6.2 kpc (see \S\ref{subsec:d} below for the distance estimate).
Since we include 4FGL J1924.3+1628 in the background model, the flux of `Src A' is a bit lower than that reported in \citet{araya21}.
The flux of 4FGL J1924.3+1628 in 0.2--500 GeV, also assumed with a PL spectrum, is $5.14\times10^{-12}\ {\rm erg\ cm^{-2} s^{-1}}$ (with 
${\rm \Gamma=1.70\pm 0.17}$).
The overall TS values of `Src A' and 4FGL J1924.3+1628 are 381.7 and 26.0, corresponding to the significance of 19.5$\sigma$ and 4.2$\sigma$, respectively.

\begin{table}
  \centering
  \caption{Formulae and free parameters for different $\gamma$-ray spectra models.}
    \begin{tabular}{lcc}
    \toprule
    \toprule
    Name & Formula & Free parameters\\
    \midrule
    PL & ${\rm d}N/{\rm d}E = N_0 (E/E_0)^{-\rm \Gamma}$ 
    & $N_0$, ${\rm \Gamma}$\\
    PLEC & ${\rm d}N/{\rm d}E = N_0 (E/E_0)^{-\rm \Gamma}{\rm exp}(-E/E_{\rm cut})$ 
    & $N_0$, ${\rm \Gamma}$, $E_{\rm cut}$\\
    LogP & ${\rm d}N/{\rm d}E = N_0 (E/E_0)^{-{\rm \Gamma} -\beta \log(E/E_0)}$ 
    & $N_0$, ${\rm \Gamma}$, $\beta$\\
    BPL & ${\rm d}N/{\rm d}E = \left\{
    \begin{array}{ll}
     {N_0 (E/E_0)^{-\rm \Gamma_1},\ E \leq E_{\rm b}} \\
     {N_0 (E/E_0)^{-\rm \Gamma_2},\ E > E_{\rm b}}
    \end{array}
    \right.$ 
    & $N_0$, ${\rm \Gamma_1}$, ${\rm \Gamma_2}$, $E_{\rm b}$\\
    \bottomrule
    \end{tabular}%
  \label{tab:fml}%
\end{table}%


\begin{table}
  \centering
  \caption{Results from spectral analysis of `Src A'.}
    \begin{tabular}{lcc}
    \toprule
    \toprule
    Spectral model & $\Delta k$ $^a$ & $\Delta$AIC $^a$\\
    \midrule
    PL & 0 & 0\\
    PLEC & 1 & 20.54\\
    LogP & 1 & 2.00\\
    BPL & 2 & 2.51\\
    \bottomrule
    \end{tabular}%
    \begin{tablenotes}
		\footnotesize
		\item \textbf{Notes.}
		\item $^a$ $k$ and AIC values are provided as differences with respect to the PL model.
	\end{tablenotes}
  \label{tab:spec_aic}%
\end{table}%

\begin{figure}
    \centering
    \includegraphics[width=\columnwidth]{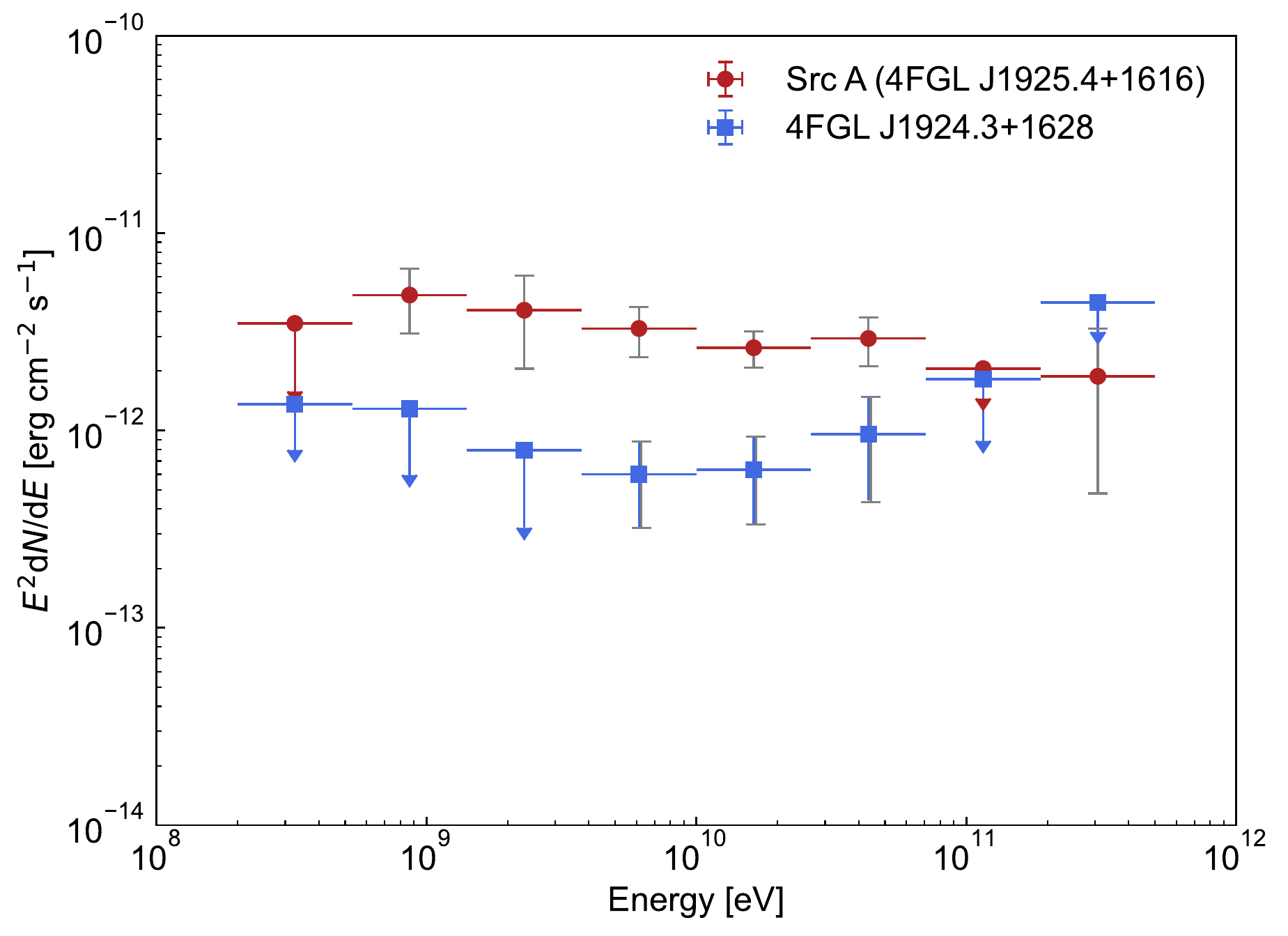} 
    \caption{
    The spectral energy distributions of `Src A' (4FGL J1925.4+1616) and 4FGL J1924.3+1628 at 200 MeV to 500 GeV.
    The combination of statistic and systematic errors are shown in gray bars.
    }
  \label{fig:sedsys}
\end{figure}

The SEDs of the two sources are extracted from the maximum likelihood analysis in eight logarithmically spaced energy bins (see Figure \ref{fig:sedsys}). 
In each bin, we fit the normalization parameters of the diffuse backgrounds and the sources within 5$^\circ$ from the ROI centre.
Additionally, we consider the systematic uncertainty caused by the model imperfection of the Galactic diffuse background.
Using an alternative Galactic diffuse emission model (\emph{gll\_iem\_v06.fits}), we repeat the above analysis to regenerate the SEDs and calculate the flux deviation as systematic errors.
If the TS value of the source in a bin is less than 4, the 95\% upper limit of its flux is estimated.

\subsection{CO Line Emission} \label{subsec:co}
\begin{figure*}
    \includegraphics[width=1.0\textwidth]{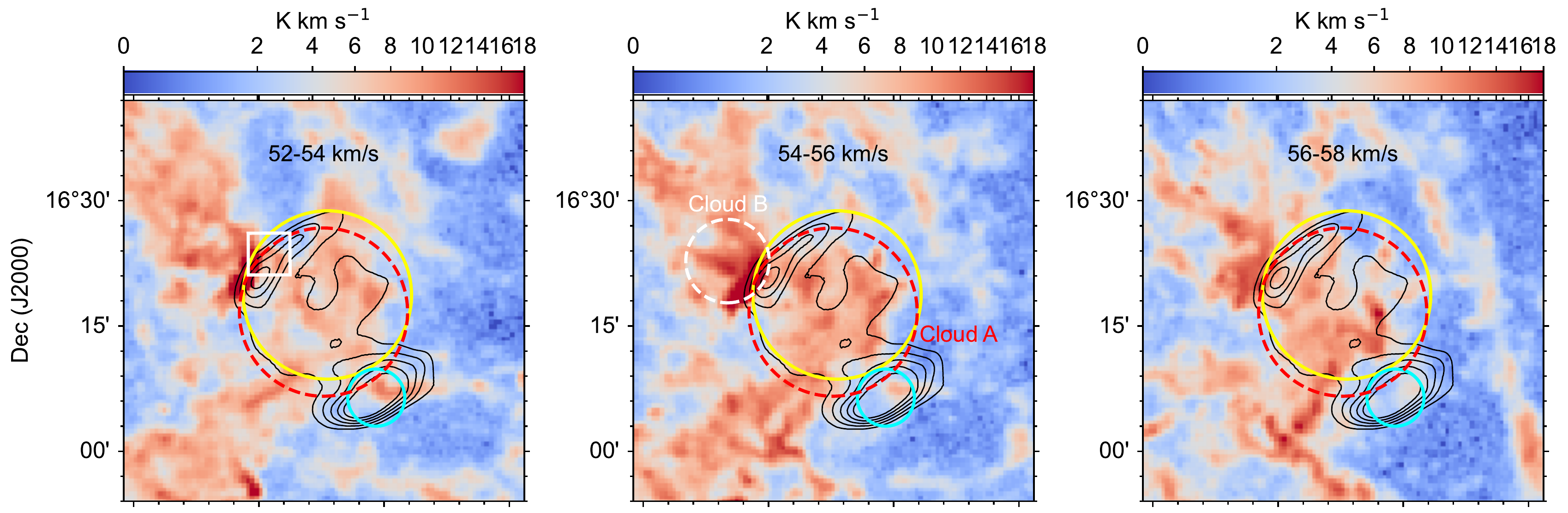} 
    \includegraphics[width=1.0\textwidth]{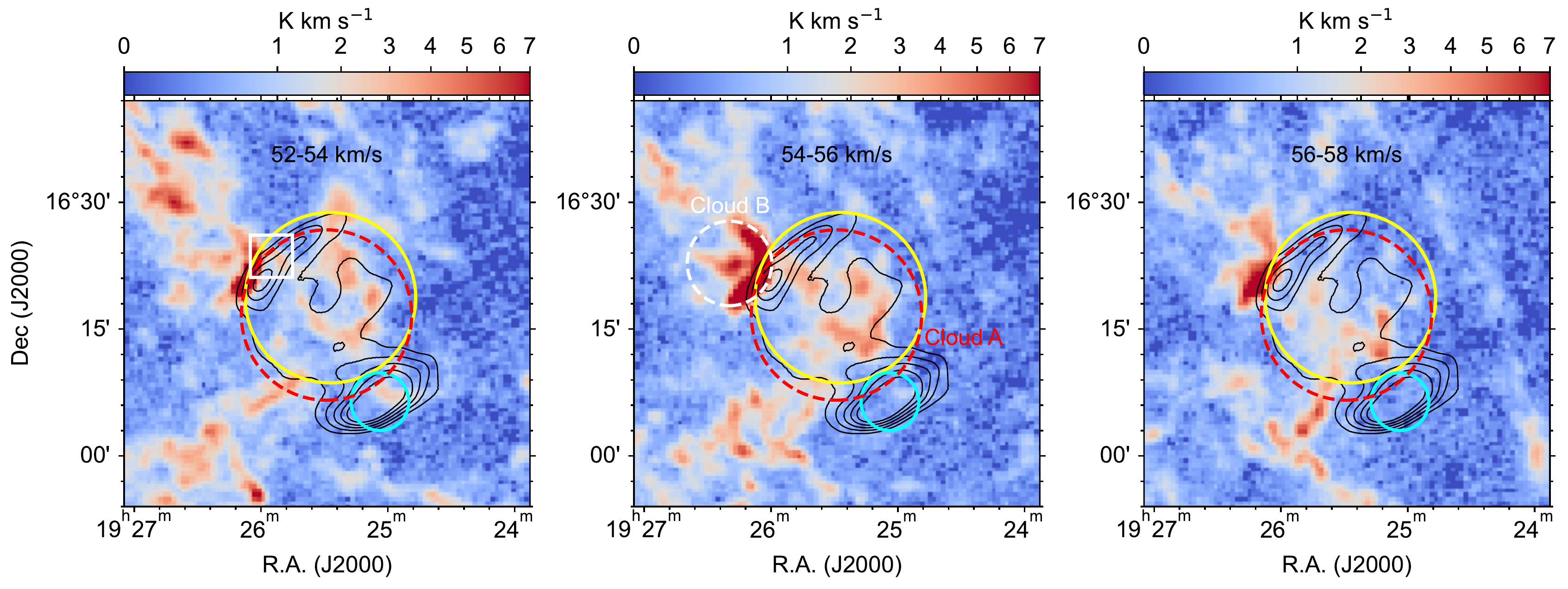}
        \caption{
        $^{12}$CO ($J=1$ -- $0$) (upper panels) and $^{13}$CO ($J=1$ -- $0$) (lower panels) intensity maps integrated each 2 km s$^{-1}$ in the velocity range of +52 -- +58 km s$^{-1}$, overlaid by GLEAM 200 MHz radio continuum emission in contours (black) with levels of 1.00, 1.25, 1.50, 1.75, and 2.00 Jy/beam.
        The yellow and cyan circles are the same as those in Figure \ref{fig:radio}.
        The red circle here indicates the diffuse cloud `Cloud A', which has the same position and size as `Src A'.
        The white box in the left panel denotes the region from which the $^{12}$CO and $^{13}$CO spectra (see Figure \ref{fig:gridmap}) are extracted.
        The dashed white circle marked as `Cloud B' in the middle panel shows the northeastern structure that has a morphological correspondence with the SNR shell.
        }
  \label{fig:co}
\end{figure*}
\begin{figure*}
    \includegraphics[width=0.8\textwidth]{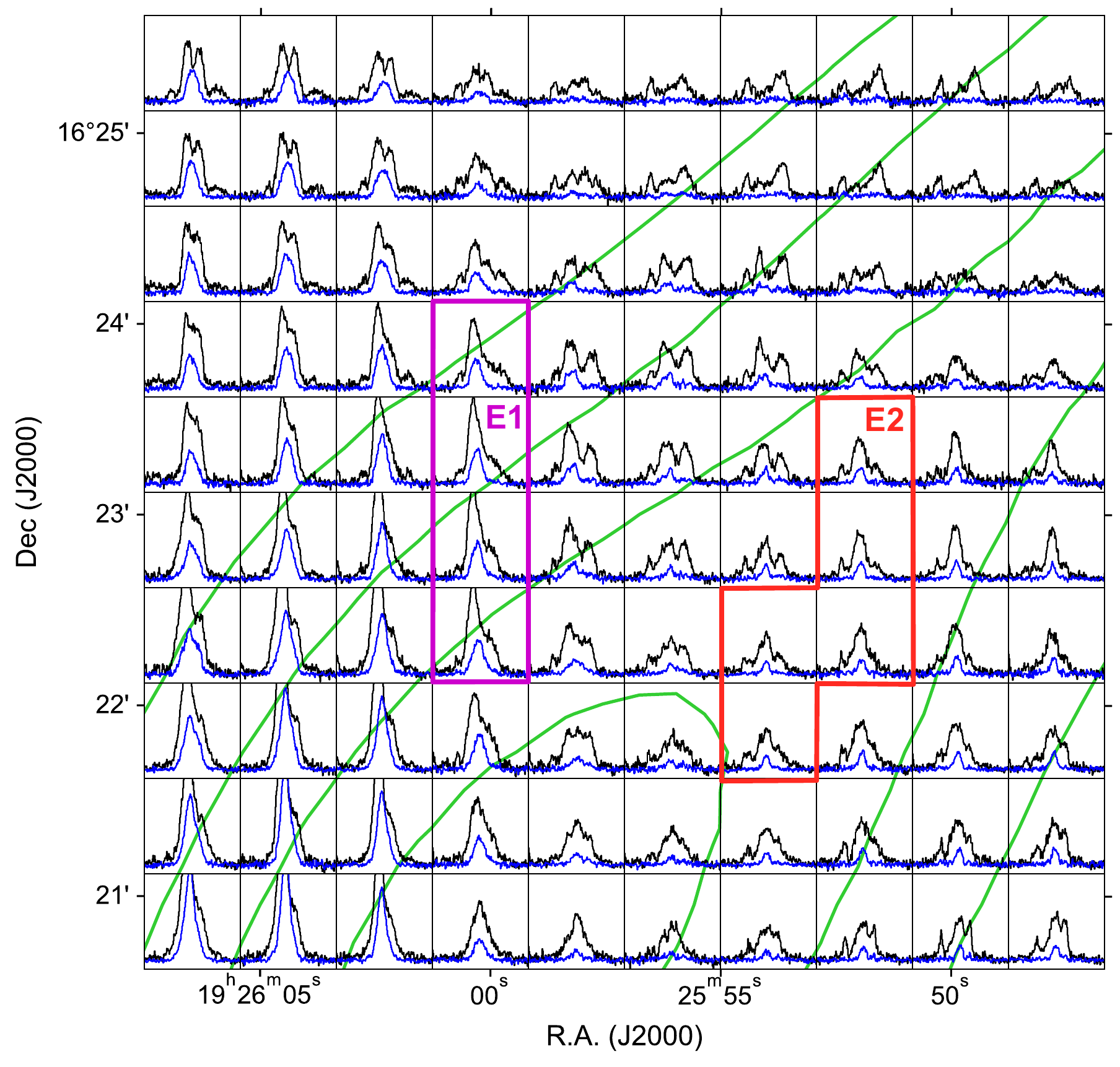} 
        \caption{
        Grid of $^{12}$CO ($J=1$ -- $0$) (black) and $^{13}$CO ($J=1$ -- $0$) (blue) spectra restricted to the velocity range +40 -- +70 km s$^{-1}$.
        Two regions, `E1' and `E2', are defined for CO-spectrum extraction (see Figure \ref{fig:mcspec}).
        The contours are the same as those in Figure \ref{fig:co}.
        }
  \label{fig:gridmap}
\end{figure*}
\begin{figure*}
    \includegraphics[width=1.0\textwidth]{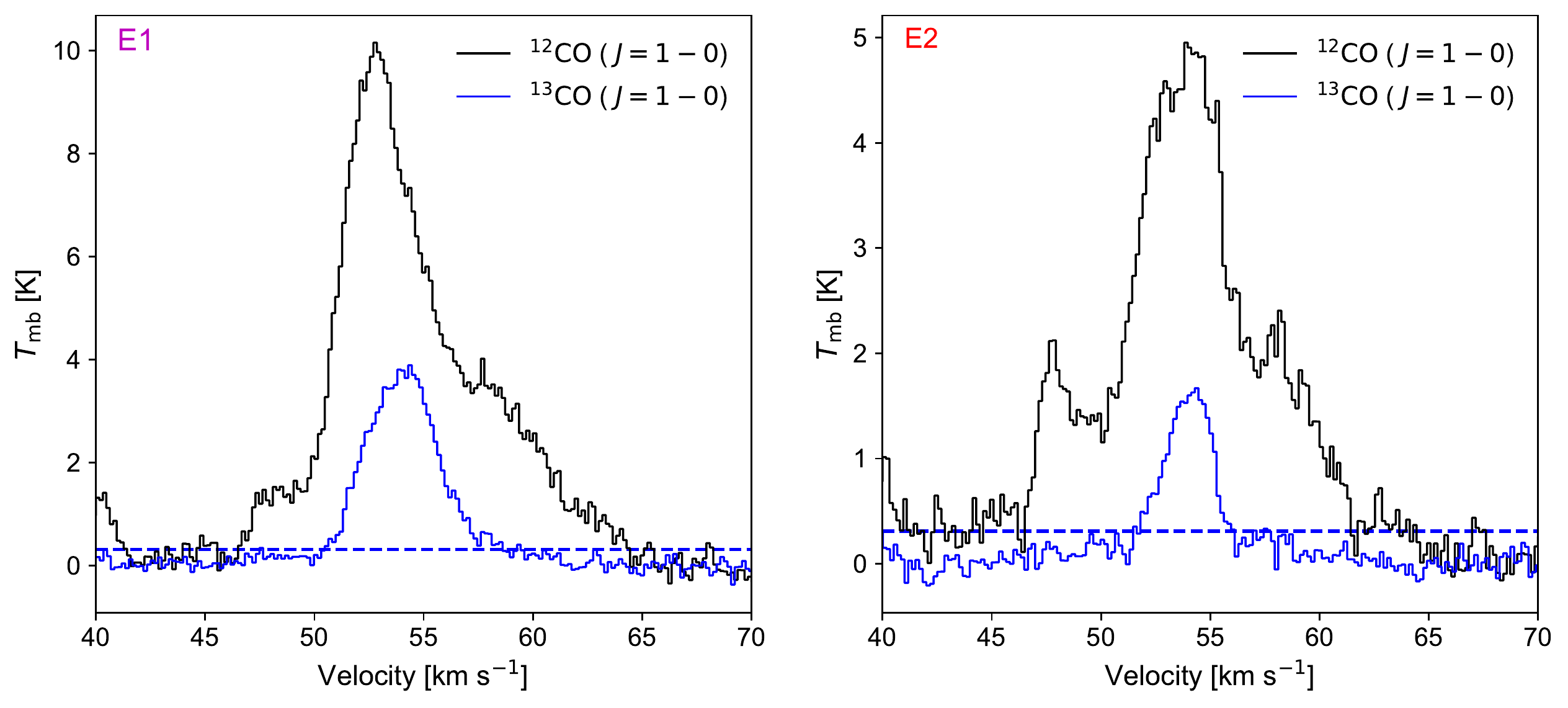} 
        \caption{
        Averaged CO line profiles of the two regions marked in Figure \ref{fig:gridmap}.
        The $^{12}$CO line profiles are plotted with black lines and the $^{13}$CO line profiles with blue lines.
        The blue dashed lines indicate the 3$\sigma$ uncertainty.
        }
  \label{fig:mcspec}
\end{figure*}
We investigate the potential molecular gas which may contribute to the hadronic emission.
Figure \ref{fig:co} shows the $^{12}$CO ($J=1$ -- $0$) and $^{13}$CO ($J=1$ -- $0$) line intensity channel maps over successive $\sim$ 2 km s$^{-1}$ intervals in the +52 -- +58 km s$^{-1}$ velocity range.
Diffuse CO gas in the interval +54 -- +56 km s$^{-1}$ seems to pervade the extent of `Src A', which is roughly coincident with SNR G51.26+0.11; we denote this part of molecular gas as `Cloud A'.
An extended, curved structure (marked as`Cloud B' in the middle panels of Figure \ref{fig:co}) in the northeast appears in contact with, and tangent to, the shell of SNR G51.26+0.11, where the radio emission is brightened, which may be indicative of the collision of the SNR shell with the cloud.
We thus generate a grid of CO-line spectra toward a $5^\prime \times 5^\prime $ region at the northeastern boundary of the SNR, marked with white boxes in the left panels of Figure \ref{fig:co}.
As shown in the grid presented in Figure \ref{fig:gridmap},
the wings of the line profiles of the CO emission at the LSR velocity $v_{\rm LSR} $ around $\sim$ +54 km s$^{-1}$ appear to be broadened in a number of pixels, as exemplified by regions `E1' and `E2'.
\\
\indent The average $^{12}$CO line profile of `E1', spanning from +45 to +65 km s$^{-1}$, seems to be asymmetrically broadened from +57 km s$^{-1}$ to +65 km s$^{-1}$ (see the left panel of Figure \ref{fig:mcspec}).
The peak velocity of $^{12}$CO emission ($\sim$ +53 km s$^{-1}$) is also slightly shifted compared to that of $^{13}$CO ($\sim$ +54 km s$^{-1}$).
Unlike the $^{12}$CO line profile of `E1', the average profile of `E2', with a velocity span similar to that of `E1', appears to be symmetrically broadened (see the right panel of Figure \ref{fig:mcspec}).
The fitted parameters for the $\sim$ +54 km s$^{-1}$ emission lines in regions `E1' and `E2' are summarized in Table \ref{tab:co_par}.
Combined with the intensity channel maps, we note that `E1' is located in the curved dense structure `Cloud B', while `E2' appears to be in the extent of `Cloud A', which overlaps the `Src A' disk.
Considering there is no significant $^{13}$CO emission (which is optically thin and represents the quiescent molecular gas) at the wings of the $\sim$ +54 km s$^{-1}$ lines, the broadened $^{12}$CO profiles suggest that both `Cloud A' and `Cloud B' are likely perturbed by the shock of SNR G51.26+0.11, and thus provide kinematic evidence for the SNR-MC interaction \citep[e.g.,][]{jiang10}.

\subsection{H\,{\small I} Absorption} \label{subsec:h1}
\begin{figure}
    \includegraphics[width=\columnwidth]{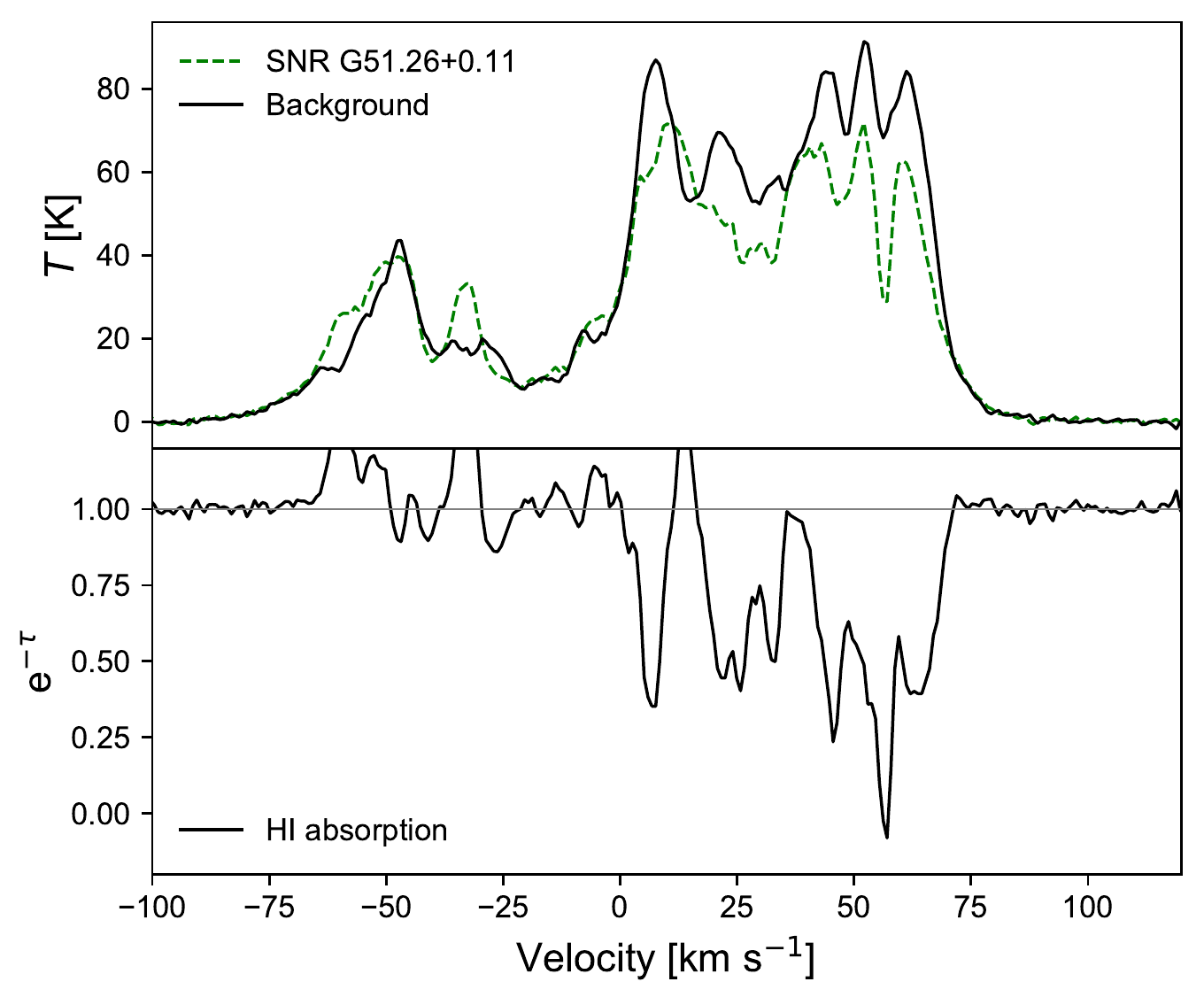} 
        \caption{
        H\,{\small I} emission (upper panel) and absorption (lower panel) spectra of SNR G51.26+0.11.
        }
  \label{fig:h1}
\end{figure}
We follow the methods of \citet{tian07} to build the 21 cm H\,{\small I} absorption spectrum.
Denote the average H\,{\small I} brightness temperatures of the on-source and the background regions at velocity $v$ as $T_{\rm on}(v)$ and $T_{\rm off}(v)$, respectively.
One then has
$T_{\rm on}(v)=T_{B}(v)(1-e^{-\tau_t(v)})+T^c_s(e^{-\tau_c(v)}-1)$ and
$T_{\rm off}(v)=T_{B}(v)(1-e^{-\tau_t(v)}+T^c_{\rm bg}(e^{-\tau_c(v)}-1)$,
where $T^c_s$ and $T^c_{\rm bg}$ are the average continuum brightness temperatures for the on-source and background regions, respectively, 
$T_{B}(v)$ refers to the spin temperature of the H\,{\small I} cloud,
$\tau_t(v)$ is the total optical depth along line-of-sight,
and $\tau_c(v)$ is the optical depth from the continuum source to the observer.
Then the absorption spectrum can be expressed as 
$e^{-\tau_c(v)}=1-({T_{\rm off}(v)-T_{\rm on}(v)})/({T^c_s-T^c_{\rm bg}})$.
The boxes, both with a size of $3^\prime \times 3^\prime $, in Figure \ref{fig:radio} show the on-source and the background regions we select.
The obtained H\,{\small I} emission and absorption spectra are displayed in Figure \ref{fig:h1}.

\section{Discussion} \label{sec:discuss}

\subsection{Molecular Environment} \label{subsec:mc}

To estimate the molecular column density and the mass of the molecular gas, local thermodynamic equilibrium for the molecular gas and optically thick conditions for the $^{12}$CO ($J=1$ -- $0$) line are assumed.
We first calculate the excitation temperature $T_{\rm ex}={5.53}/{\ln[1+{5.53}/({T_{\rm peak}(^{12}{\rm CO})+0.819})]}$ K \citep{naga98},
where $T_{\rm peak}(^{12}{\rm CO})$ is the peak temperature of the maximum $^{12}$CO($J=1$ -- $0$) emission point in the cloud region.
The peak excitation temperature of the dense gas in `Cloud B' is $T_{\rm ex}\sim 24$ K, which is noticeably higher than the typical temperature of $\sim$ 10 K in quiescent interstellar MCs.
Then we estimate the column density via 
$N({\rm H}_2) =1.49\times10^{20}\ {W(^{13}{\rm CO})}/[{1-\exp (-5.29/T_{\rm ex})}]$ cm$^{-2}$ \citep{naga98},
where $W$($^{13}$CO) is the integrated intensity of the $^{13}$CO ($J=1$ -- $0$) line.
We take the abundance ratio [H$_2$]/[$^{13}$CO] as $7\times10^5$ \citep{frerk82}.
After estimating the $N({\rm H}_2)$ value in each pixel of the corresponding regions, we obtain the average values for the clouds.
The relation $M=2.8m_{\rm H}N({\rm H}_2)A$ is used to obtain the mass of the molecular gas,
where $m_{\rm H}$ is the mass of a hydrogen atom and $A$ is the cross-sectional area of the corresponding clouds.
The clouds are simply approximated as spheres when calculating the average number densities of hydrogen nuclei ($n_{\rm H}$). 
The parameters of `Cloud A' and `Cloud B' are summarized in Table \ref{tab:mc}.

\begin{table}
  \centering
  \caption{Fitted parameters for region `E1' and `E2'.}
    \begin{tabular}{lccccc}
    \toprule
    \toprule
    \multirow{2}[2]{*}{Region} & \multirow{2}[2]{*}{Line} & Centre & FWHM & $T_{\rm peak}$ & $W$\\
      &   & (km s$^{-1}$) & (km s$^{-1}$) & (K) & (K km s$^{-1}$)\\
    \midrule
    \multirow{3}[2]{*}{E1} & \multirow{2}[1]{*}{$^{12}$CO ($J=1-0$) $^a$} & 52.8 & 3.0 & 6.5 & \multirow{2}[2]{*}{66.4}\\
      &   & 55.4 & 10.0 & 4.2 &  \\
      & $^{13}$CO ($J=1-0$) & 54.1 & 4.0 & 3.8 & 17.2\\
    \midrule
    \multirow{3}[2]{*}{E2} & \multirow{2}[1]{*}{$^{12}$CO ($J=1-0$) $^a$} & 53.9 & 3.3 & 2.8 & \multirow{2}[2]{*}{39.5}\\
      &   & 53.9 & 12.9 & 2.2 & \\
      & $^{13}$CO ($J=1-0$) & 54.1 & 2.8 & 1.6 & 6.0\\
    \bottomrule
    \end{tabular}%
    \begin{tablenotes}
		\footnotesize
		\item \textbf{Notes.}
		\item $^a$ For $^{12}$CO, two Gaussian models are used to fit the spectra.
	\end{tablenotes}
\label{tab:co_par}%
\end{table}%


\begin{table*}
  \centering
  \caption{Properties of the molecular clouds `Cloud A' and `Cloud B'.}
    \begin{tabular}{lcccccccc}
    \toprule
    \toprule
    \multirow{2}[2]{*}{MC} & (R.A., Dec) $^a$ & $R$ & $d$ & $N$(H$_2$) & $M$ & $n_{\rm H}$ & ${T_{\rm ex}}^b$ & $\tau$($^{13}$CO) $^c$\\
      & ($^{\circ}$) & ($^{\circ}$) & (kpc) & (10$^{21}$ cm$^{-2}$) & (10$^4d^2{\rm M}_{\odot}$) & (cm$^{-3}$) & (K) & \\
    \midrule
    Cloud A & (291.38, 16.27) & 0.17 & 6.2 & 2.8 & 6.4 & 70 & 13.4 & 0.31\\
    \midrule
    Cloud B & (291.59, 16.38) & 0.08 & 6.2 & 8.0 & 4.6 & 430 & 24.5 & 0.86\\
    \bottomrule
    \end{tabular}%
    \begin{tablenotes}
		\footnotesize
		\item \textbf{Notes.}
		\item $^a$ Centre of the clouds.
		\item $^b$ The excitation temperature of CO.
		\item $^c$ The optical depth of $^{13}$CO: $\tau$($^{13}$CO) $\approx -\ln[1-T_{\rm peak}(^{13}{\rm CO})/T_{\rm peak}(^{12}{\rm CO})]$.
	\end{tablenotes}
  \label{tab:mc}%
\end{table*}%


\subsection{Kinematic Distance} \label{subsec:d}

Considering the association with the molecular clouds seen at +43.4 -- +52.2 and +53 -- +58.8 km s$^{-1}$, SNR G51.26+0.11 was placed at a distance between 3.7 and 6.6 kpc \citep{araya21}.
The distance was also estimated to be $6.6\pm1.7$ kpc from the H\,{\small I} absorption spectra that were constructed with the THOR data \citep{ranas22}.
Based on the association with the $\sim+54$ km s$^{-1}$ MCs, we here can provide an independent estimate for the kinematic distance of the SNR.
Applying the Galactic rotation curve together with $R_0$ = 8.34 kpc and $V_0$ = 240 km s$^{-1}$ \citep{reid14}, the systemic LSR velocity of +54 km s$^{-1}$ corresponds to a near distance $4.2\pm0.5$ kpc and a far distance $6.2\pm0.5$ kpc \citep{wenger18}.
Additionally, the predicted tangent point distance toward SNR G51.26+0.11 is at $\sim$ 5.2 kpc and the tangent point LSR velocity is $\sim$ +52 km s$^{-1}$.
As shown in Figure \ref{fig:h1}, the highest H\,{\small I} emission velocity appears at $\sim$ +62 km s$^{-1}$.
Whether the tangent point LSR velocity is $\sim$ +52 km s$^{-1}$ or $\sim$ +62 km s$^{-1}$, continuous H\,{\small I} absorption features are present from +45 km s$^{-1}$ to the tangent point.
This suggests that SNR G51.26+0.11 is located beyond the tangent point, and indicates that the SNR and the associated MCs are most probably at the far kinematic distance, $6.2\pm0.5$ kpc.
The distance we determine here is within the ranges given by the previous studies.

\subsection{Origin of the Gamma-ray Emission} \label{subsec:origin}
\subsubsection{Pulsar Origin} \label{subsubsec:psr}
We first discuss the possibility that the $\gamma$-ray emission comes from pulsars.
As shown in Figure \ref{fig:tsmap}, three pulsars, PSR J1926+1613, PSR J1924+1631, and PSR J1924+1628, are seen to be located within the ROI when we search in the Australia Telescope National Facility (ATNF; \citealt{manch05})\footnote{\url{https://www.atnf.csiro.au/research/pulsar/psrcat/}} 
pulsar catalogue and the SIMBAD Astronomical Database \citep{wenger00}\footnote{\url{http://simbad.u-strasbg.fr/simbad/sim-fcoo}}.
PSR J1926+1613 is projected inside the disk template for `Src A' and near the edge of the SNR shell.
Since the distance ($d$) and the spin-down luminosity of PSR J1926+1613 are unknown, we can not determine the contribution from the pulsar.
However, the extended GeV morphology and flat SED of `Src A' seem unfavorable to a pure emission from the pulsar or the putative PWN.
For 4FGL J1924.3+1628, both PSR J1924+1631 and PSR J1924+1628 are much closer (with an angular distance of $\sim 0.08^{\circ}$ and $\sim 0.13^{\circ}$, respectively) to it than the SNR is.
According to the dispersion measures, PSR J1924+1631 and PSR J1924+1628 are predicted to be at distances of 14.0 kpc \citep{nice13} and 10.5 kpc \citep{psf22}, respectively.
For 4FGL J1924.3+1628, it would have a luminosity of $\sim 5\times10^{34}(d/10.5\ {\rm kpc})^2$ erg s$^{-1}$ in the energy range 0.2--500 GeV, which is two orders of magnitude higher than the spin-down luminosities of $2.4\times10^{32}$ erg s$^{-1}$ for PSR J1924+1628 and $5.7\times10^{32}$ erg s$^{-1}$ for PSR J1924+1631. 
Thus these two pulsars seem incapable of powering the $\gamma$-ray source 4FGL J1924.3+1628.

\subsubsection{SNR G51.26+0.11} \label{subsubsec:snr}
\begin{figure*}
    \centering
    \includegraphics[width=1.0\textwidth]{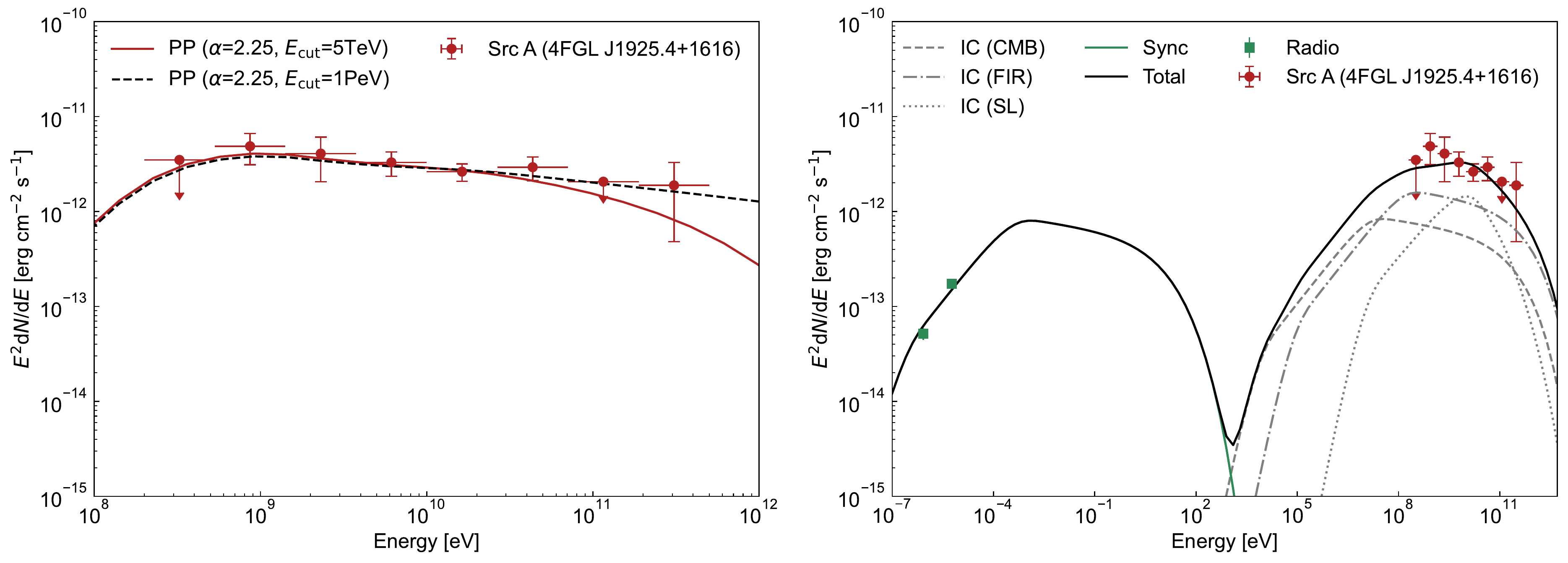} 
        \caption{
        The fitted hadronic (left) and leptonic (right) models for the SED of `Src A' (4FGL J1925.4+1616).\
        The radio data of SNR G51.26+0.11 are taken from \citet{dokara21}.
        Details of the models are described in Section \ref{subsubsec:snr}.
        }
  \label{fig:sedA}
\end{figure*}
Noting that `Src A' essentially coincides SNR G51.26+0.11 in the line of sight, we focus on the SNR origin of `Src A'.
As shown above, there is also a spatial coincidence between the SNR and the +54 km s$^{-1}$ MCs at a distance of $\sim$ 6.2 kpc.
Moreover, the broadened profiles of the CO emission lines indicate that the MCs, `Cloud A' and `Cloud B', both are likely perturbed by the shock of SNR G51.26+0.11.
Thus a plausible explanation for `Src A' is the hadronic interaction between the relativistic protons accelerated by the shock of SNR G51.26+0.11 and the adjacent MCs.
To analyze the possible contribution from SNR G51.26+0.11 to the $\gamma$-ray emission of `Src A', we fit the obtained $\gamma$-ray spectrum in both hadronic and leptonic scenarios using the {\small Naima} package (version 0.9.1, \citealt{zabz15}).
We assume the accelerated protons and electrons have a power-law distribution in energy with an exponential cutoff $E_{i,\rm cut}$:
\begin{equation}
    {\rm d}N_{i}/{\rm d}E_{i} \propto E^{-\alpha_{i}} {\rm exp}(-E_{i}/E_{i,\rm cut}),
\end{equation}
where $i={\rm p, e}$, $E_{i}$ is the particle kinetic energy, and $\alpha_{i}$ is the power-law index.
The normalization is determined by the total energy in particles with energies above 1 GeV ($W_{i}$).
\\
\indent In the hadronic scenario, by adopting a distance of 6.2 kpc to the SNR, the proton index $\alpha_{\rm p} \approx 2.25$ and the lower limit of the proton cutoff energy $E_{\rm p,cut}\approx$ 5 TeV are obtained with the GeV data alone (see the left panel of Figure \ref{fig:sedA}).
The proton index ($\alpha_{\rm p}$) we obtain is slightly larger than that (2.11$^{+0.08}_{-0.12}$) reported by \citet{araya21} and hence more consistent with the known gamma-ray-bright interacting SNRs \citep{acero16},
and a PWN contribution is not necessary to explain the hard spectrum given \emph{1Disk+1PS} as the best template.
The cutoff energy can not be well constrained and is fixed as 1 PeV in the calculation.
We first consider the contribution from the SNR interaction with `Cloud A', which contains all the diffuse molecular gas in the disk template of `Src A'.
Because `Cloud A' almost overlaps with the SNR along the line of sight, the fraction of the SNR-MC interaction region to the SNR surface is assumed to be $f\sim0.5$.
With the parameters displayed in Table \ref{tab:mc}, we use $n_{\rm H}\sim70$ cm$^{-3}$ as the number density of the target gas for proton-proton (PP) hadronic interaction.
The energy required in protons is 
$W_{\rm p} \sim 2.0\times10^{49}$($d$/6.2 kpc)$^2$($n_{\rm H}$/70 cm$^{-3}$)$^{-1} (f/0.5)^{-1}$ erg, 
which is feasible in the context of SNR scenario.
For comparison, we also estimate the contribution from the SNR interaction with `Cloud B' in the northeast.
The interaction region, assumed as a spherical cap, represents a fraction of the SNR surface with an order of $f\sim 0.01$.
We set the value of $n_{\rm H}$ to be 430 cm$^{-3}$ and obtain
$W_{\rm p} \sim 1.6\times 10^{50}$($d$/6.2 kpc)$^2$($n_{\rm H}$/430 cm$^{-3}$)$^{-1} (f/0.01)^{-1}$ erg, 
which is also acceptable for the SNR scenario.
But it is likely that the molecular gas surrounding the SNR, including `Cloud A' and `Cloud B', all contribute to the hadronic emission \footnote{It cannot be ruled out that `Cloud A' and `Cloud B' are comprised in the same MC complex. Although no significant spectral line-broadening features can be distinguished in the central regions of `Cloud A', this does not imply that no interaction occurs.}.
As the $\gamma$-ray flux $F \propto \eta E_{\rm SN} f n_{\rm H}$, where $\eta$ is the acceleration efficiency and $E_{\rm SN}$ is the explosion energy of the supernova, the ratio between the $\gamma$-ray emission from the two clouds, i.e., `Cloud A' and `Cloud B', is estimated to be $\sim 8.14$.
Thus the central diffuse gas contributes up to 90\% of the observed $\gamma$-ray emission, satisfactorily explaining the morphology of `Src A'.
On the other hand, if the molecular gas is proximate to the SNR, even if it is not directly shocked, the accelerated protons that escape from the shock front may still hit the molecules \citep[e.g.,][]{li&chen10, ohira11, li&chen12}, giving rise to hadronic $\gamma$-rays.
Despite the total energy requirements would be increased in this case, the canonical supernova explosion energy is sufficient in our estimation.
\\
\indent For the leptonic scenario, we assume the radio and $\gamma$-ray emission are generated by the same electron population.
Their interaction with the magnetic field, i.e., the synchrotron (Sync) radiation, accounts for the radio emission and the inverse Compton (IC) scattering of low energy photons accounts for the $\gamma$-ray emission.
Incorporating the radio fluxes of G51.26+0.11 taken from \citet{dokara21}, we fit the broad-band SED using an exponential cutoff broken power-law with a break energy $E_{\rm break}$ and a cutoff energy $E_{\rm e,cut}$.
Due to synchrotron losses of the electrons, we set the spectral index below the break to be $\alpha_{\rm e}$ and above the break to be $(\alpha_{\rm e}$+1).
Lack of constraint by the current data, the cutoff energy is fixed 10 TeV based on the X-ray observations on the shell-type SNRs \citep{reynolds99}.
Besides the cosmic microwave background (CMB), the seed photon fields we use include a far-infrared (FIR) field with a temperature of 30 K and an energy density of 0.5 eV cm$^{-3}$ and a star light (SL) field with a temperature of 4000 K and an energy density of 1.0 eV cm$^{-3}$, estimated at a distance of 6.2 kpc from the interstellar radiation field model \citep{shi11}.
As shown in the right panel of Figure \ref{fig:sedA}, the data can be fitted with parameters $\alpha_{\rm e} \approx 2.1$ and $E_{\rm break}\approx 50$ GeV, as well as magnetic field $B \approx 3.2\mu$G.
The total required energy in primary electrons is $W_{\rm e} \sim 3.4\times10^{49}$($d$/6.2 kpc)$^2$ erg,
which would imply that the leptonic scenario is also feasible energetically.
However, the cooling time is about $10^7$ yr to produce the break energy of 50 GeV in the magnetic field of 3.2 $\mu$G, which seems to be unphysically long for an SNR with an obvious radio shell morphology.
Considering the break feature cannot be explained by the cooling losses alone,
alternative models, such as age-limited \citep{ohira17, zhang19} or escape-limited \citep{ohira10, ohira12, hess18} acceleration, should be included.
This is beyond the scope of this study and not investigated here.
In addition, the low magnetic field strength obtained is somewhat lower than expected in the diffusive shock acceleration paradigm.
As the scenario is also subject to constraints from X-ray emission, relevant observations of SNR G51.26+0.11 are encouraged.

\section{Conclusions} \label{sec:sum}

We reanalyze the \emph{Fermi}-LAT $\gamma$-ray emission in the region of SNR G51.26+0.11 and study its interstellar environment based on the CO-line and H\,{\small I} data.
Below are the main results of this study:

\begin{enumerate}[i.]
\item Using the MWISP survey data, we find some MCs at $v_{\rm LSR}$ $\sim$ +54 km s$^{-1}$ overlap with SNR G51.26+0.11 along the line of sight and a dense structure tangent to the northeastern shell of the SNR.
Given the morphological coincidences between the molecular gas distribution and the SNR and the asymmetric or broad CO line profiles, it is very likely that the MCs are associated with the SNR.
\item According to the LSR velocity of the associated MCs and the H\,{\small I} absorption features, SNR G51.26+0.11 is found to be located behind the tangent point at a farther distance, $6.2\pm0.5$ kpc.
\item Based on 13.2 years of \emph{Fermi}-LAT observations, 
we resolve the extended $\gamma$-ray emission detected in the region of SNR G51.26+0.11 into a uniform-disk $\gamma$-ray source (`Src A'/4FGL J1925.4+1616) with a significance of about 19.5$\sigma$ and a point source (4FGL J1924.3+1628) with a significance of about 4.2$\sigma$ in 0.2--500 GeV.
The best-fitting position and 68\%-containment radius of `Src A' are R.A. = $291.38\pm 0.02^{\circ}$, Dec = $16.27\pm 0.02^{\circ}$ (J2000) and $\sigma={0.17^{+0.02}_{-0.02}}^{\circ}$.
Its emission can be described by a power-law spectrum with a photon index of ${\rm \Gamma=2.19\pm 0.04}$.
The corresponding 0.2--500 GeV flux is $2.86\times10^{-11}\ {\rm erg\ cm^{-2}\ s^{-1}}$,
and the luminosity in 1--100 GeV is $7.3\times10^{34}$ erg s$^{-1}$ at a distance of 6.2 kpc.
\item A natural explanation for `Src A' is the hadronic scenario in which the energetic protons accelerated by SNR G51.26+0.11 interact with the adjacent MCs.
The proton spectrum obtained has an index of $\sim$ 2.25.
The leptonic scenario, even if unlikely, cannot be conclusively ruled out.
\end{enumerate}

\section*{Acknowledgements}

This research made use of the data from the Milky Way Imaging Scroll Painting (MWISP) project, 
which is a multi-line survey in 12CO/13CO/C18O along the northern galactic plane with PMO-13.7m telescope. We are grateful to all the members of the MWISP working group for their support.
W.J.Z.\ thanks Bing Liu, Qiancheng Liu, Tianyu Tu, Ping Zhou and Xin Zhou for helpful discussions.
X.Z.\ and Y.C.\ acknowledge the support of National Key R\&D Program of China under nos.\ 2018YFA0404204 and 2017YFA0402600 and the NSFC grants under nos.\ U1931204, 12173018, and 12121003.

\section*{Data Availability}

The \emph{Fermi}-LAT and VLA H\,{\small I} data underlying this work are publicly available, and can be downloaded from 
\url{https://fermi.gsfc.nasa.gov/ssc/data/access/lat/}
and
\url{https://www.cadc-ccda.hia-iha.nrc-cnrc.gc.ca/en/search/}, respectively.
The CO data used in this study will be made available by the corresponding authors upon request.




\bibliographystyle{mnras}
\bibliography{g51} 







\bsp	
\label{lastpage}
\end{document}